\documentclass[preprint,12pt,numbers,square,numbers]{elsarticle}
\usepackage{amssymb}
\usepackage{array}
\newcolumntype{P}[1]{>{\centering\arraybackslash}p{#1}}
\newcolumntype{M}[1]{>{\centering\arraybackslash}m{#1}}
\usepackage{a4wide,amssymb,epsfig,latexsym,multicol,array,hhline,fancyhdr,pictex, latexsym, graphicx,amsbsy,amsfonts,amsthm,verbatim}
\usepackage{supertabular}
\usepackage{amsmath}
\usepackage{lastpage}
\usepackage[lined,boxed,commentsnumbered]{algorithm2e}
\usepackage{enumerate}
\usepackage{color}
\usepackage{longtable}
\usepackage{arydshln}
\usepackage{graphicx}
\usepackage{array}
\usepackage{booktabs}
\usepackage{indentfirst}
\usepackage{tabularx, caption}
\usepackage{multirow}
\usepackage{multicol}
\usepackage{rotating}
\usepackage{graphics}
\usepackage{setspace}
\usepackage{epsfig}
\usepackage{tikz}
\usepackage{pgfplots}
\usepackage{array}
\usepackage{float}
\usepackage{wrapfig}
\usepackage{indentfirst}
\usepackage{tikz-timing}[2014/10/29]
\usetikztiminglibrary[rising arrows]{clockarrows}
\usepackage{xparse}
\usetikzlibrary{arrows,backgrounds}
\usepackage{hyperref}
\usetikzlibrary{arrows,shapes.gates.logic.US,shapes.gates.logic.IEC,calc}
\hypersetup{urlcolor=blue,linkcolor=black,citecolor=black,colorlinks=true} 
\usepackage{footnote}
\usetikzlibrary{patterns}
\usepackage[]{algorithm2e}
\usepackage{xcolor}
\usepackage{listings}
\usepackage{color}
\usepackage{soul}
\usepackage{layout}
\usepackage{subcaption} % For subfigures
\usepackage[pass,a4paper,margin=1in,bindingoffset=0.5in]{geometry}
\usepackage{tabularray}
\pgfplotsset{compat=1.18}
\newcommand{\PreserveBackslash}[1]{\let\temp=\\#1\let\\=\temp}
\newcolumntype{C}[1]{>{\PreserveBackslash\centering}p{#1}}
\newcolumntype{R}[1]{>{\PreserveBackslash\raggedleft}p{#1}}
\newcolumntype{L}[1]{>{\PreserveBackslash\raggedright}p{#1}}
\usepackage{listings}
\usepackage{color}

\definecolor{dkgreen}{rgb}{0,0.6,0}
\definecolor{gray}{rgb}{0.5,0.5,0.5}
\definecolor{mauve}{rgb}{0.58,0,0.82}

\definecolor{dkgreen}{rgb}{0,0.6,0}
\definecolor{gray}{rgb}{0.5,0.5,0.5}
\definecolor{mauve}{rgb}{01,0,0.82}
\lstdefinelanguage{Python}{
    keywords={def, return},
    keywordstyle=\color{blue},
    commentstyle=\color{gray},
    stringstyle=\color{green},
    identifierstyle=\color{black},
    basicstyle=\ttfamily\small,
    breaklines=true,
    showstringspaces=false,
    numbers=left,
    numberstyle=\tiny,
    numbersep=5pt,
    frame=tb,
    frameround=tttt,
    tabsize=4,
    columns=flexible,
    keepspaces=true
}

\usepackage{xcolor,pifont}
\newcommand*\colourcheck[1]{%
  \expandafter\newcommand\csname #1check\endcsname{\textcolor{#1}{\ding{51}}}%
}
\colourcheck{blue}
\colourcheck{green}
\colourcheck{red}

\NewDocumentCommand{\busref}{som}{\texttt{%
#3%
\IfValueTF{#2}{[#2]}{}%
\IfBooleanTF{#1}{\#}{}%
}}

\definecolor{darkgray}{rgb}{.4,.4,.4}
\definecolor{purple}{rgb}{0.65, 0.12, 0.82}

\lstdefinelanguage{js}{
  keywords={div, Button, img, Logo, Link, style},
  keywordstyle=\color{black}\bfseries,
  ndkeywords={class, export, boolean, throw, implements, import, const},
  ndkeywordstyle=\color{black}\bfseries,
  identifierstyle=\color{black},
  sensitive=false,
  comment=[l]{//},
  morecomment=[s]{/*}{*/},
  stringstyle=\color{black}\ttfamily,
  morestring=[b]',
  morestring=[b]"
}

\usetikzlibrary{arrows,backgrounds}
\usepackage{hyperref}
\hypersetup{urlcolor=blue,linkcolor=black,citecolor=black,colorlinks=true}

\begin{document}

\begin{frontmatter}

% \title{FM-based-ACG: An FM-based automatic code generation system with parameter optimization}
\title{CodeLSI: Leveraging Foundation Models for Automated Code Generation with Low-Rank Optimization and Domain-Specific Instruction Tuning}
\author[inst1,inst2]{Huy Le}
\author[inst1,inst2]{Phong Nguyen}
\author[inst1,inst2]{Hao Do}
\author[inst1,inst2]{Tuan Nguyen}
\author[inst1,inst2]{Thien Pham}
\author[inst3]{Anh Nguyen-Duc}
\author[inst1,inst2]{Tho Quan}

\affiliation[inst1]{organization={Faculty of Computer Science and Engineering, Ho Chi Minh City University of Technology (HCMUT),  Vietnam}}
\affiliation[inst2]{organization={Vietnam National University Ho Chi Minh City,  Vietnam}}
\affiliation[inst3]{organization={Department of Economics and IT, University of South Eastern Norway, Norway}}

\begin{abstract}
Context: Automated code generation using Foundation Models (FMs) offers promising solutions for enhancing software development efficiency. However, challenges remain in ensuring domain specificity, cost-effectiveness, and security—especially when relying on third-party APIs. This paper introduces CodeLSI, a framework that combines low-rank optimization and domain-specific instruction tuning to address these challenges.

Objectives: The aim of this study is to develop and evaluate CodeLSI, a novel approach for generating high-quality code tailored to specific domains, using FMs fine-tuned on company infrastructure without dependence on external APIs.

Methods: CodeLSI applies low-rank adaptation techniques to reduce the computational cost of model pre-training and fine-tuning. Domain-specific instruction tuning is employed to align code generation with organizational needs. We implemented and tested the framework on real-world JavaScript coding tasks using datasets drawn from internal software projects.

Results: Experimental evaluations show that CodeLSI produces high-quality, context-aware code. It outperforms baseline models in terms of relevance, accuracy, and domain fit. The use of low-rank optimization significantly reduced resource requirements, enabling scalable training on company-owned infrastructure.

Conclusion: CodeLSI demonstrates that combining low-rank optimization with domain-specific tuning can enhance the practicality and performance of FMs for automated code generation. This approach provides a secure, cost-efficient alternative to commercial API-based solutions and supports faster, more targeted innovation in software development.
\end{abstract}


\end{frontmatter}

\newpage

\section{Introduction}
\label{sec:introductiontotopic}
%\textit{Automatic code generation} \citep{75}  is a significant focus within both Artificial Intelligence and Software Development fields. Its aim is to produce programming code automatically based on user-provided requirements in natural language. It promises to simplify programming, reduce coding time, and assist novice programmers in their learning process. Several authors argue that this is the $"$holy grail$"$ of Computer Science \citep{manna1971toward} \citep{gulwani2017program} due to its complexity and significance when achieved.  Traditional approaches like \emph{Model-driven architecture} \citep{soley2000model} \citep{mdarchi1} or \emph{Template-based code generation} \citep{syriani2018systematic}  \citep{cao2020survey} have limitations, often requiring rigid templates and struggling with uncovered requests. Recent advancements in Artificial Intelligence, especially in natural language understanding \citep{radford2019language}, offer promising solutions. Modern AI models leverage vast datasets, including billions of code snippets, to comprehend and process user requests effectively. Moreover, to generate code tailored to specific domain requirements, models need training on similar projects' source code. However, privacy concerns prevent software companies from sharing their code for third-party training. Hence, organizations prefer to develop their own domain-specific code generation models. 

Automatic code generation has long been regarded as a critical and aspirational goal in Software Engineering research \citep{budinsky_automatic_1996,herrington_code_2003,dehaerne_code_2022}. Its core objective is to translate user-provided requirements—often expressed in natural language—into executable source code, thereby streamlining the software development process. This capability promises to reduce development effort, accelerate prototyping, and support less experienced developers in learning programming practices. The significance of this challenge has led some to refer to it as the "holy grail" of computer science \citep{manna1971toward, gulwani2017program}.

Traditional code generation techniques, such as model-driven architecture (MDA) \citep{soley2000model, mdarchi1} and template-based generation methods \citep{syriani2018systematic, cao2020survey}, offer some degree of automation but are often constrained by predefined structures and lack flexibility when faced with novel or domain-specific requirements. These limitations have spurred interest in AI-driven approaches, particularly with the advent of large-scale Foundation models (FMs), which those trained on vast corpora of source code and natural language, have introduced new possibilities for flexible and intelligent code generation \citep{radford2019language}.

Current research revealed challenges when applying these FM-based code generation approaches in real-world software project contexts \citep{liu_exploring_2024,wang_review_2023,jiang_survey_2024,joel_survey_2024}. Notably, high-quality code generation in specific application domains requires training on representative, domain-specific codebases. Yet, due to intellectual property concerns and data privacy regulations, many software organizations are reluctant to share proprietary code, hindering the ability to fine-tune general models for specific industrial needs. Hence, there remains a clear lack of effective, open, and adaptable code generation tools for other widely-used languages such as TypeScript. This gap presents significant challenges for Web application developers and organizations with domain-specific needs. While general-purpose FMs such as Claudie and Gemini can generate code in multiple languages, they often lack fine-grained domain knowledge and cannot be easily customized due to proprietary constraints. Moreover, fine-tuning large-scale models with hundreds of billions of parameters is computationally intensive and cost-prohibitive for many organizations \citep{chen2024exploring, lialin2023scaling}.

To address these limitations, we propose and empirically evaluate a lightweight and cost-efficient method for automated code generation using Low-Rank Adaptation (LoRA) \citep{hu2021lora} in combination with Instruction Tuning \citep{wei2022finetuned, longpre2023flan}. These techniques are integrated into a domain-specialized model, CodeLSI, designed to support TypeScript-based applications and multiple downstream software engineering tasks, including code generation and summarization. The research is guided by the following research questions:
\begin{itemize}
    \item RQ1. How does integrating LoRA and Instruction Tuning affect FM performance in code generation? 
    \item RQ2. Can the proposed approach meet domain-specific needs, particularly for TypeScript based applications? 
    \item RQ3. What are the cost and resource benefits compared to traditional FM training? 
\end{itemize}
By answering these questions, this paper aims to advance automated code generation and offer practical solutions for businesses.

% ===========
%\clearpage

%\begin{figure}[htb!]
%        \centering
%        \includegraphics[scale=0.225]{images/code_logo_explanation.png}
%        \caption[Result of code explanation for logo creation problem using a general knowledge model]{Result of code explanation for logo creation problem using a general knowledge model.}
%        \label{fig:code_block_introduction_result}
%\end{figure}

% With the power of combining downstream tasks by adapting Instruction tuning, CodeLSI is also capable of giving a detail explanation for the code generated as illustrated in Figure \ref{fig:code_block_introduction_result}.

 % \clearpage

This paper endeavors to address the challenging problem of code generation and proposes an automated framework for TypeScript and other programming languages, catering to both generic code generation and focusing on domain-specific project requirements. Our contributions are outlined as follows:
\begin{itemize}
    \item We created a 20,000-pair instruction dataset—one of the first large-scale, publicly available resources tailored to the TypeScript language.
    \item We developed a domain-specific dataset from real-world projects to train CodeLSI models beyond the scope of mainstream code generation tools.
    \item We introduced CodeLSI, an efficient auto-generated code model combining FMs, LoRA-based fine-tuning, and prompt tuning, supporting both code generation and summarization.
    \item We built a flexible framework for integrating FMs with language-specific fine-tuning strategies across specialized datasets.
\end{itemize}
   
The study is organized as follows: Secion 2 presents the study background, Section 3 proposes our model CodeLSI, Section 4 are two experiments with the model evaluation, Section 5 discusses the result and Section 6 concludes the paper.
% \section{Preliminaries}
\section{Background}
\label{sec:theoreticalfoundations}
\subsection{Common Terminology}
% \label{sec:FM}
% Large Language Models (FMs) are Artificial Intelligence systems trained to understand and process natural language \citep{zhao2024explainability} \citep{chang2024survey}. They are a type of machine learning model that uses very deep artificial neural networks and can handle a large amount of language data. An FM can be trained on a vast dataset containing millions of sentences and words. The training process involves predicting the next word in a sentence based on the previous ones, enabling the model to "learn" the grammar and syntax of the language \cite{ozdemir2023quick}. With powerful computational abilities and numerous parameters, these models can learn complex relationships in language and generate natural sentences. Current large language models have made significant progress in natural language processing tasks, including machine translation, headline generation, question answering, and automatic text generation. \citep{khurana2023natural} They can produce high-quality natural text closely matching the content and context provided to them. 

% Recently, a common approach to improving model performance has been increasing the size of parameters and training data, leading to remarkable leaps in machines' natural language processing capabilities.
In the rapidly evolving field of Artificial Intelligence (AI), numerous specialized terms are frequently used. Understanding these terms is crucial for comprehending the concepts and discussions in this paper. This section provides definitions and references for key terms such as Natural Language Processing (NLP), Generative AI (GenAI), Foundation Model, Fine-Tuning, Prompt Engineering, and Bidirectional Encoder Representations from Transformers (BERT). The following table lists these terms along with their definitions and references.

\begin{table}[H]
\centering
\begin{tabular}{|c|p{2.5cm}|p{10cm}|p{2cm}|}
\hline
\textbf{Id} & \textbf{Term} & \textbf{Definition} & \textbf{Reference} \\
\hline
1 & NLP & Natural Language Processing (NLP) is a branch of AI that focuses on the interaction between computers and humans through natural language. & \cite{manning1999foundations} \\
\hline
2 & GenAI & Generative AI (GenAI) refers to AI systems capable of generating text, images, or other media in response to prompts. & \cite{goodfellow2014generative} \\
\hline
3 & FMs & FMs are deep learning models with a large number of parameters, designed to understand and generate human-like text. & \cite{devlin2018bert} \\
\hline
4 & Fine-Tuning & Fine-Tuning is the process of taking a pre-trained model and further training it on a specific task or dataset to improve performance. & \cite{howard2018universal} \\
\hline
5 & Prompt Engineering & Prompt Engineering involves designing and refining prompts to elicit the desired response from AI models, particularly in NLP tasks. & \cite{brown2020language} \\
\hline
6 & BERT & Bidirectional Encoder Representations from Transformers (BERT) is a transformer-based model designed for understanding the context of words in a sentence. & \cite{devlin2018bert} \\
\hline
\end{tabular}
\caption{Glossary of Common Terminology}
\label{tab:glossary}
\end{table}

\subsection{Low-Rank Adaptation}
In October 2018, the BERT model \citep{devlin2019bert}, with 350 million parameters, marked a breakthrough in Transformer models. However, even modern hardware at the time struggled to fine-tune such large models due to memory constraints. Since then, model sizes have exploded—reaching up to 540 billion parameters \citep{chowdhery2022palm}—far outpacing the growth in computational resources and making full fine-tuning impractical for most users.

Despite this, FMs have demonstrated strong generalization capabilities across diverse tasks, even those outside their original training domains. This has led to efficient adaptation techniques like parameter-efficient fine-tuning, which aim to retain performance while minimizing training costs. One prominent method is Low-Rank Adaptation (LoRA) \citep{hu2021lora, gu2024mix}, which freezes the original model weights and introduces a small number of trainable parameters through low-rank matrix decomposition. This approach significantly reduces computational overhead while leveraging the knowledge embedded in the original model. The transformation matrix can be represented as follows:
\[
\delta W = W_A W_B
\]
\[
W_A \in \mathbf{R}^{\mathrm{in}\times r}, W_B \in \mathbf{R}^{\mathrm{out}\times r}
\]

All parameters of the pre-trained model are frozen, and only the matrices \(W_A\) and \(W_B\) are trainable. The scaling factor is a constant and is usually set to \(1/r\). After training, they can be integrated into the original matrix \(W\) by adding the \(W_A W_B\) matrix to the original \(W\) matrix. In Transformers, LoRA is often applied to the \(W_K\) and \(W_V\) matrices in multi-head attention layers. This method has been evaluated on models with up to 175 billion parameters and has shown superior results compared to other parameter tuning methods (BitFit, Adapter). It even outperforms traditional fine-tuning (FT) methods that fine-tune all original weight matrices while using only 2 \% of the parameter count

\begin{figure}[htb!]
        \centering
        \includegraphics[scale=0.7]{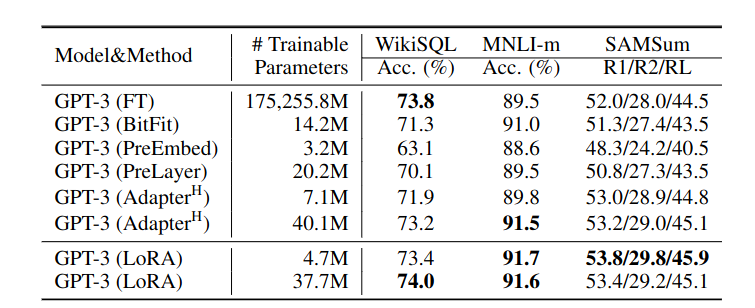}
        \caption[Comparison of Performance between LoRA and other methods on a 175-billion-parameter GPT-3 model]{Comparison of Performance between LoRA and other methods on a 175-billion-parameter GPT-3 model \citep{hu2021lora}.}
        % \label{tab:benchmark}
\end{figure}

% \newpage
\subsection{Instruction Tuning Method}
\label{sec:instruction_tuning}

Instruction tuning \citep{pmlr-v202-longpre23a} is proposed as an effective technique to enhance the capabilities and controllability of large language models. It involves further training large language models using (INSTRUCTION, OUTPUT) pairs, where INSTRUCTION represents human instructions for the model, and OUTPUT represents the desired output following the INSTRUCTION. The aim is to align the FM's predictions with specific human instructions, thereby bridging the gap between the inherent next-word prediction objective of FMs and the user's goal of instruction following. By utilizing it, FMs exhibit more controllable and predictable behaviors compared to standard models. The instructions provided during tuning serve as constraints that guide the model's outputs towards desired response characteristics or domain-specific knowledge. This not only enhances model controllability but also provides a means for humans to intervene and influence the model's behaviors effectively. Moreover, it offers computational efficiency by enabling rapid adaptation of FMs to specific domains without the need for extensive retraining or significant architectural changes.

\begin{figure}[!htb]
    \centering
    \includegraphics[scale=0.45]{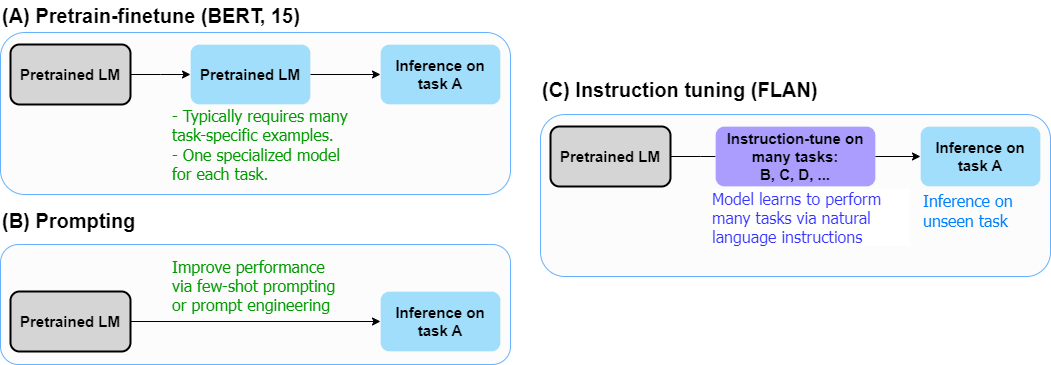}
    \caption[Comparison between instruction tuning and two types of request-based fine-tuning and pre-training fine-tuning]{Comparison between instruction tuning and two types of request-based fine-tuning and pre-training fine-tuning \citep{wei2022finetuned}.}
    \label{fig:lialin2023scaling2}
\end{figure}

Instruction tuning is also extremely useful for handling many downstream tasks by synthesizing them into the initial (INSTRUCTION, OUTPUT) pair. Therefore, the code generation system, which was originally divided into two steps (code generation and code summarization), can now combine these processes in its output. Moreover, it is highly suitable for handling domain-specific datasets as mentioned earlier. Therefore, we have decided to use this tuning method in our CodeLSI system.

% \section{Related Works}
%\textit{In this section, we provide a general overview of the related research works that have been and are being conducted, as well as the current trends in solving the problem. We also discuss and evaluate these methods as they form an important foundation for their research during the paper process.}

\label{sec:relatedresearch}
\subsection{Classical approaches for automatic code generation}
\label{sec:traditional-approaches}
The problem of automatic code generation has been around for over 50 years, dating back to the latter half of the last century \citep{75}, and continues to evolve over time. The approach of that era mostly utilized \textit{template-based} code generation methods \citep{syriani2018systematic}, with the idea of inputting specific data into predefined templates to create new code segments.

\begin{figure}[htb!]
        \centering
        \includegraphics[scale=0.46]{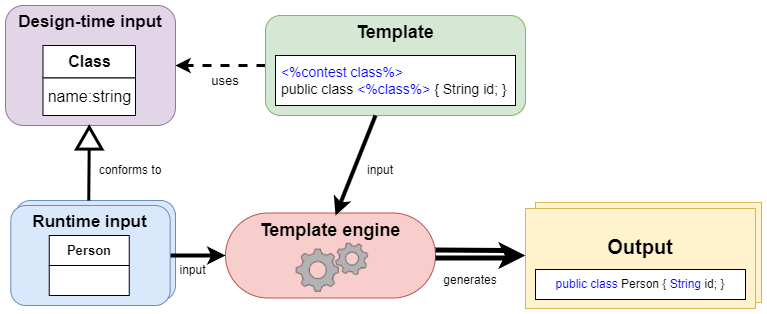}
        \caption[Components of the template-based code generation model]{Components of the template-based code generation model \citep{syriani2018systematic}.}
        \label{fig:tbcg}
\end{figure}

Template-based code generation relies on predefined templates to produce code outputs. Each template consists of a static part that remains constant in the output and a dynamic part, or pseudo code, which the system generates at runtime. A code generator executes these templates based on runtime inputs to replace pseudo code with specific code segments. The design-time inputs specify the constraints and information guiding the code generation process to ensure appropriate code segments are produced. While this method is straightforward and suitable for predefined scenarios, it has limitations. Firstly, it offers limited flexibility as code segments are constrained by predefined templates, making customization challenging if templates do not cover all potential variations. Secondly, defining and maintaining templates, especially for complex codebases, can be complex and error-prone, leading to challenges in understanding, updating, and debugging. Lastly, the approach can introduce development challenges, such as consistency issues when updating multiple code segments generated from a single template or synchronizing changes between templates and already generated code segments. Despite efforts in the early 2000s to enhance flexibility by generating code based on queries \citep{soley2000model}, these approaches faced challenges in dealing with program complexity and accurately capturing user requirements \citep{manna1971toward} \citep{gulwani2017program}. Recent studies in the 2010s focused on addressing these challenges using methods like stochastic search \citep{schkufza2013stochastic} and deductive top-down search based on user queries  \citep{gulwani2011automating} \citep{polozov2015flashmeta}, yet scalability remained an issue. The ability to accurately recognize and express user intentions, whether through logical rules, input-output examples, or natural language, remains a key challenge that traditional approaches have yet to fully solve

\subsection{FM-based leveraged approaches for automatic code generation}
\label{sec:modern-approaches}
Foundation model (FM)-based architectures, particularly transformer-based models such as GPT (Generative Pre-trained Transformer), have opened new opportunities to train machines that can understand and generate code based on diverse, real-world user requirements \citep{bengesi2024advancements}. These models are trained on large-scale textual corpora and source code datasets, enabling them to learn linguistic patterns, programming syntax, and semantic structures through unsupervised pre-training. Fine-tuning these models on domain-specific or task-specific data further improves their performance in software engineering tasks such as code completion, generation, and summarization. By adapting pre-trained models with relatively small datasets, researchers and practitioners can tailor general-purpose models to domain-specific applications with significantly reduced computational cost.

Over the past few years, industry leaders have introduced a series of high-capacity large language models (FMs) that have redefined the state of the art in AI-driven code generation and natural language understanding. Notable examples include OpenAI’s GPT-4 and GPT-4 Turbo, Google’s Gemini 1.5 Pro, and DeepMind’s AlphaCode 2 and Chinchilla. These models have demonstrated exceptional performance across a wide range of software engineering tasks. However, they remain proprietary, with access restricted to controlled APIs and governed by commercial licensing or geographical limitations.

In response, the research community has accelerated efforts to develop competitive open-source FMs. Recent releases such as Meta’s LLaMA 3, Mistral’s Mixtral, HuggingFace’s StarCoder2, Stability AI’s StableLM, and TII’s Falcon 180B represent significant milestones in this space. These models have enabled broader experimentation, reproducibility, and community-driven advancement. While open models have historically lagged behind their closed-source counterparts in benchmark performance, the gap is rapidly narrowing due to innovations in model architecture, fine-tuning techniques, and community contributions \citep{kukreja2024literature, cai_demystifying_2025,ahmed_studying_2024}.

To address the scalability challenges associated with fine-tuning large models, recent research has focused on parameter-efficient fine-tuning (PEFT) techniques. These methods aim to minimize the number of trainable parameters while preserving model performance, making fine-tuning feasible even with limited resources. Techniques such as Low-Rank Adaptation (LoRA), adapters, and prompt tuning have become popular in adapting large models to specific domains or tasks. Empirical studies, such as “Scaling Down to Scale Up” \citep{lialin2023scaling}, have benchmarked these methods, highlighting trade-offs in memory efficiency, performance, and training cost. Instruction tuning is another promising improvement as combines pre-trained models with curated instructional datasets to improve their ability to follow user commands expressed in natural language \citep{wei2022finetuned, longpre2023flan}.  

In the latter half of 2023 alone, major releases included HuggingFace's StarCoder under the BigCode initiative \citep{li2023starcoder}, Meta’s LLaMA 2, and Google’s PaLM-2 and Gemini, integrated into Bard. Alongside these developments, performance optimization techniques such as DeepSpeed ZeRO-3 for distributed training \citep{aminabadi2022deepspeed} and PagedAttention for faster inference \citep{kwon2023efficient} are pushing the boundaries of scalability and efficiency in FM deployment.

These trends collectively highlight a vibrant and rapidly evolving field. However, for software engineering practice, the empirical evaluation of these technologies—especially in the context of domain-specific tasks, programming languages like TypeScript, and cost-efficient training strategies—remains underexplored. This gap motivates the need for rigorous, practice-oriented research that investigates the application of FMs in realistic development environments.

\section{The LSI Code Generation Model}
\label{sec:proposedmodel}

% \subsection{Implementation and Experimental Results}
We implemented the proposed code generation model by developing two main models: the \emph{CodeLSI General Knowledge Model} and the \emph{CodeLSI Specialized Knowledge Model} for TypeScript language.

\subsection{CodeLSI General Knowledge Model}

\begin{figure}[!htb]
        \centering
        \includegraphics[scale=0.47]{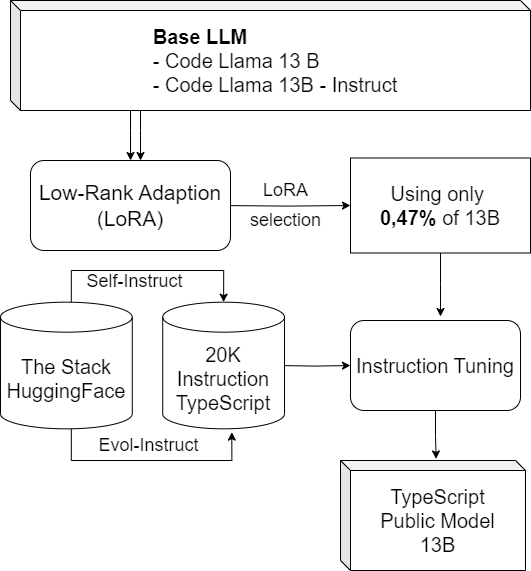}
        \caption[CodeLSI General Knowledge Model Architecture]{CodeLSI General Knowledge Model Architecture}
        \label{fig:architecture-overview}
\end{figure}

The CodeLSI General Knowledge model focuses on automated code generation using general knowledge from the TypeScript-Instruct 20K dataset. This knowledge is derived from open-source code snippets across various domains, enabling our initial model to generate diverse code for any general user requirement. We enhanced two variants of a base large language model with 13 billion parameters, which was already capable of automated code generation but not specifically tailored to TypeScript. To better align the model with TypeScript code generation, we aimed to fine-tune the base model and use this refined version as a foundation for subsequent models discussed in this paper. We chose to implement Low-Rank Adaptation (LoRA) and Instruction Tuning for this purpose. As mentioned, LoRA reduces the number of parameters that need fine-tuning while still delivering excellent results compared to retraining all 13 billion parameters. In our case, we used LoRA to train only 0.47\% of the parameters. Instruction Tuning was applied using {instruction, output} data pairs. Ultimately, this led to the creation of our initial model, the CodeLSI General Knowledge model. The detailed process flow of this model is defined as follows:

\begin{itemize}
   \item[\textbf{Step 1:}] Research and utilize the current best-performing large language models with an appropriate number of parameters for the task of automated code generation. In this paper, we proposed Code Llama model with two variants Code Llama - 13B and Code Llama - Instruct 13B
   \item[\textbf{Step 2:}] Apply parameter optimization methods with instruction tuning on the custom-built TypeScript-Instruct 20K dataset, derived from the larger The Stack dataset. We use the LoRA method, the Insutruction Tuning technique, and the DeepSpeed tool \citep{aminabadi2022deepspeed} to accelerate training, resulting in the general-purpose CodeLSI General Knowledge.
   \item[\textbf{Step 3:}] The CodeLSI General Knowledge 13B can automatically generate TypeScript code based on natural language requests from users. This model addresses the primary objective of the thesis. We evaluate this model on the Multi-Lingual Human Eval dataset for TypeScript  \citep{cassano2022multiple} and tests it with general coding requests.
\end{itemize}

\subsection{CodeLSI Specialized Knowledge Model}
\label{sec:model-2}
\begin{figure}[!htb]
        \centering
        \includegraphics[scale=0.454]{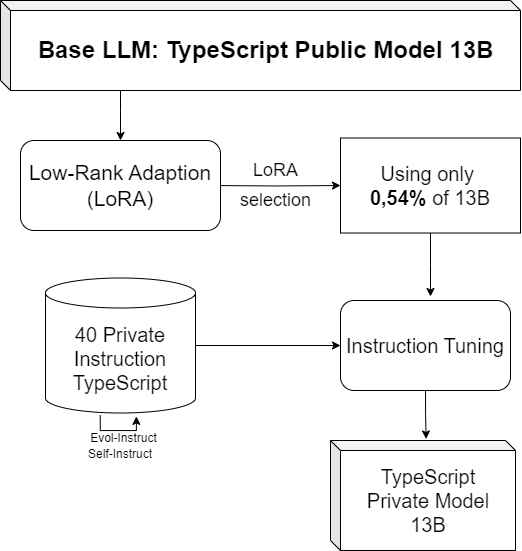}
        \caption[CodeLSI Specialized Knowledge Model Architecture]{CodeLSI Specialized Knowledge Model Architecture}
        \label{fig:architecture-overview-2}
\end{figure}

Our initial goal was to develop a code generation model tailored for specific domain requirements. After successfully refining and building the CodeLSI General Knowledge model, we used this improved model as the base large language model to further refine the CodeLSI Specialized Knowledge model. Following the same architecture as the initial model, we applied parameter tuning methods, specifically Low-Rank Adaptation (LoRA) and instruction tuning, on a manually curated dataset of 40 private TypeScript instructions, as detailed in the Dataset section. The result was a successful TypeScript code generation model that addresses specialized private knowledge, outperforming even powerful modern tools like ChatGPT, as will be shown in the comparative results in the next chapter. The detailed process flow of this model is as follows:

\begin{itemize}
   \item[\textbf{Step 1:}]  Use our CodeLSI General Knowledge model as the base model. This clarifies the motivation behind building our initial model and helps focus the tuning process more specifically on the TypeScript language, rather than using a general-purpose model like Code Llama.
   \item[\textbf{Step 2:}] Apply parameter optimization methods with instruction tuning on the custom-built 40 Private Instruction dataset. We utilize the LoRA method, the instruction tuning technique, and the DeepSpeed tool \citep{aminabadi2022deepspeed} to accelerate training, resulting in the domain-specific CodeLSI Specialized Knowledge model.
   \item[\textbf{Step 3:}] The CodeLSI Specialized Knowledge model can automatically generate TypeScript code based on natural language user requests. Additionally, it excels in generating code for specific personal projects we have built. We tested it with domain-specific coding requests and compared its performance to initial general models to highlight its superiority in real-world projects.
\end{itemize}

\subsection{TypeScript-Instruct 20K Dataset}
\label{sec:20k}
While Python remains the dominant programming language in the development and evaluation of large language models (FMs) for code generation—reflected in tools such as HumanEval \citep{chen2021evaluating}—there is limited support for other widely used programming languages, particularly TypeScript, which is extensively employed in modern web development. A review of publicly available code-focused FMs revealed only three models that offer TypeScript support: Code Llama (Meta AI), StarCoder (Hugging Face), and DeepSeek Coder (DeepSeek). Of these, only StarCoder provides transparency regarding its training data through the BigCode project, while the training data for the other models remains undisclosed.

In response to this gap, we constructed a new dataset, TypeScript-Instruct 20K, to support the development and fine-tuning of TypeScript-capable code generation models. This dataset is inspired by the Alpaca 20K dataset—originally designed for Python—and aims to extend instruction-tuning capabilities to the TypeScript domain.

The TypeScript-Instruct 20K dataset was derived from The Stack \citep{kocetkov2022stack}, a 6TB multilingual source code corpus released as part of the BigCode Project, which includes code in over 350 programming languages. From this, we extracted TypeScript files using the pre-filtered subset the-stack-smol-xl, available through Hugging Face. We randomly sampled 10,000 TypeScript code snippets, totaling approximately 131 GB of raw data.

To convert raw code into a usable instruction dataset, we applied a multi-step data generation pipeline inspired by Self-Instruct \citep{wang2023selfinstruct}. Using OpenAI’s GPT-3.5-turbo API, we generated natural language instructions corresponding to each code snippet. To improve diversity and task complexity, we further applied the Evol-Instruct method \citep{xu2023wizardlm}, resulting in two instruction-output pairs per code snippet, yielding a total of 20,000 data points.

Data post-processing included quality filtering to remove ill-formed examples and truncation of code outputs exceeding 4096 tokens to ensure compatibility with models such as Code Llama \citep{rozière2023code}. The final dataset—TypeScript-Instruct 20K—consists of 20,000 high-quality instruction-code pairs, specifically curated for instruction tuning and fine-tuning of FMs in the TypeScript programming language.

\begin{figure}[htb!]
        \centering
        \includegraphics[scale=0.5]{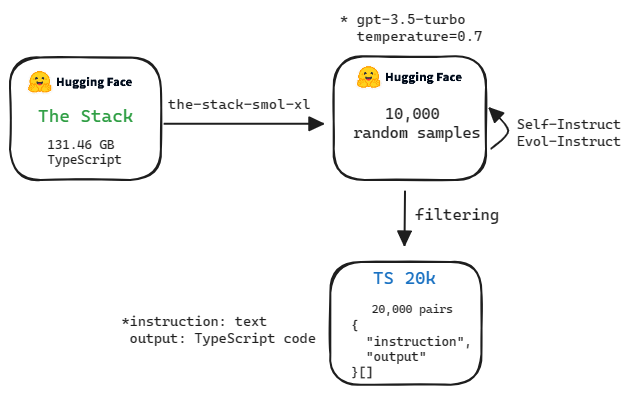}
        \caption[Processing flow for building the TypeScript-Instruct 20K dataset]{Processing flow for building the TypeScript-Instruct 20K dataset.}
        \label{fig:20k}
\end{figure}

A summary of the data generation workflow is illustrated in Figure \ref{fig:20k}.

% \newpage

\begin{lstlisting}[caption={A sample data from the dataset TypeScript-Instruct 20K. }]
Instruction: Generate the code that uses the `useStoreMap` function from the `effector-react` library to render the `foo` value from a store object. Explain how the code works.

Output: 

import React from 'react';
import { createStore } from 'effector';
import { useStoreMap } from 'effector-react';

const store = createStore({ foo: 0, bar: 'ok' });
const Foo = () => {
const value = useStoreMap(store, ({ foo }) => foo);

return <div>{value}</div>;
};

// Explanation: The `useStoreMap` function is used to create a reactive component that renders the `foo` value from the `store` object. Instead of using an object with `store`, `keys`, and `fn` properties, the `store` object itself is passed as the first argument. The second argument is a callback function that receives the current store value and returns the desired value to render. In this case, the `foo` property is extracted from the store object and rendered inside a `<div>` element.
\end{lstlisting}

\subsection{Personal Dataset}
\label{sec:40}
We proceeded to develop a simple project using TypeScript as the foundation for building and evaluating this code generation system. We want to emphasize that this is a project that has been used in the real world and has closed-source code that has not been publicly released. This helps us avoid common code generation tools that could potentially locate and train on these specific code segments, ensuring transparency in future comparisons.

\begin{figure}[!htb]
        \centering
        \includegraphics[scale=0.6]{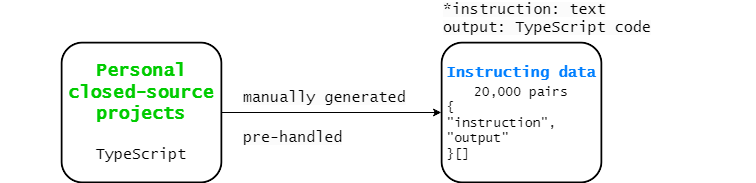}
        \caption[Processing flow for building the personal-specific dataset]{Processing flow for building the personal specific dataset.}
        \label{fig:20kb}
\end{figure}

The structure of the sample data is similar to the TypeScript-Instruct 20K dataset described earlier. It consists of instructions containing code requirements, outputs containing specific private code snippets, and explanations of the code. The data is written in TypeScript and was manually extracted by hand.

\newpage
\label{code:prepare_private_project}
\begin{lstlisting}[caption={A sample data from the Personal Dataset }, label="prepare_private_project"]
Instruction:  Create a collection that stores affiliates of shopee feature, that need an endpoint and expiration limit

Output: 
import { buildCollection } from "firecms";
import { Roles } from "../consts/auth.consts";

export const moneyCollection = buildCollection<Money>({
  name: "Money",
  singularName: "Money",
  path: "money",
  icon: "paid",
  group: "Data",
  permissions: ({ authController }) => {
    const isAdmin = authController.extra?.roles.includes(Roles.ADMIN);
    return {
      edit: isAdmin,
      create: isAdmin,
      delete: isAdmin,
      read: isAdmin,
    };
  },
  properties: {
    endpoint: {
      name: "Endpoint",
      dataType: "string",
    },
    expiration: {
      name: "Expiration",
      dataType: "number",
    },
  },
});


\end{lstlisting}

The code snippet provided above exists within the source code of our personal project. We use it to create a collection to store information about Shopee Affiliate. This collection includes a field called "Endpoint" which holds the URL leading to the sales shop online (Shopee platform) whenever a user clicks, along with the affiliate cookie. It also includes a field called "Expiration" to renew the cookie whenever it expires.
\section{Experiments}
This section first presents the base model we developed from (Section 5.1), our experiments (Section 5.2 and Section 5.3). We adopt an experimental research strategy in accordance with the principles outlined by Wohlin et al. \citep{wohlin_experiments_2024}, aiming to evaluate the effects of parameter-efficient fine-tuning methods on large language models for code generation. Specifically, we employ a quasi-experiment to compare the performance of baseline and fine-tuned models under controlled conditions \citep{kitchenham_preliminary_2002}. By systematically varying the tuning configurations (e.g., LoRA, instruction tuning) and assessing outcomes using standardized benchmarks and real-world project tasks, we are able to isolate and measure the impact of our intervention. Although randomization was not feasible due to computational constraints, we apply replication and baseline comparisons to enhance the internal validity of the findings.
\subsection{Base Model}
\label{sec:basemodel}
\label{sec:codellama}

We use Code Llama as the base model for our experiments. Released by Meta AI in August 2023, Code Llama is built on the Llama 2 architecture and is designed for code generation from natural language inputs. It is available in 7B, 13B, and 34B parameter versions, all trained on 500 billion tokens with code-related data. While larger models offer higher accuracy, the smaller 7B and 13B variants provide faster responses, making them suitable for lightweight tasks.

Two key variants are used in this study: Code Llama Python, fine-tuned on 100B Python-specific tokens, and Code Llama Instruct, which is trained via instruction tuning to follow user prompts across multiple programming languages.

\begin{figure}[!htb]
        \centering
        \includegraphics[scale=0.65]{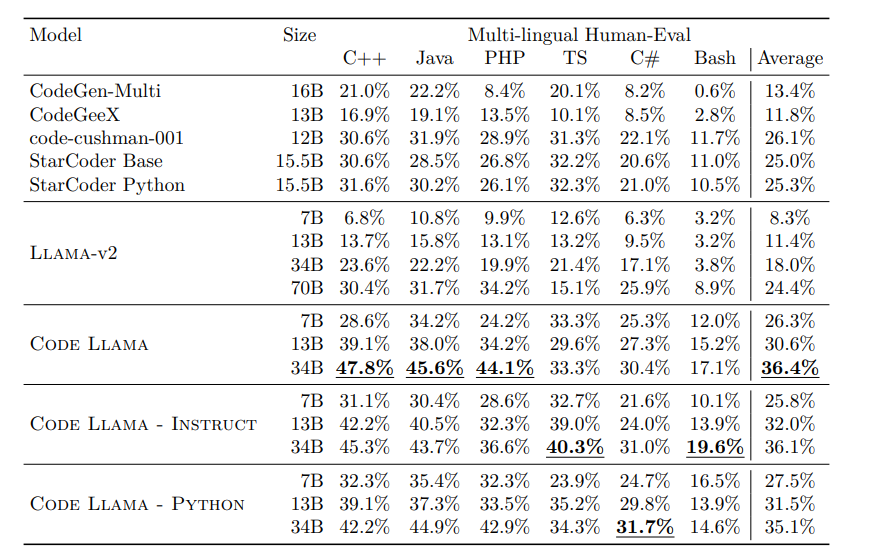}
        \caption[Evaluation results on CodeLlama models]{Evaluation results on CodeLlama models on the Multi-Lingual HE, Pass@1 scores, zero-shot \citep{rozière2023code}.}
\end{figure}

The Code Llama - Instruct model has demonstrated strong performance in code generation tasks, including for the TypeScript language, and is recognized as one of the state-of-the-art models in this domain \citep{rozière2023code}. Accordingly, we adopt Code Llama - Instruct 13B (13 billion parameters) as the baseline model for our experiments, against which we evaluate subsequent enhancements.

\subsection{Experiment 1 - Code LSI General Knowledge Model}
\subsubsection{Objective}
The objective of this experiment is to evaluate the performance of the CodeLSI General Knowledge Model, a TypeScript-focused large language model (FM) fine-tuned using the Low-Rank Adaptation (LoRA) method. We aim to compare its code generation capabilities against baseline and state-of-the-art models, particularly in terms of accuracy and efficiency, using established benchmarks for source code generation.
\subsubsection{Experiment setup}
We use Code Llama - Instruct 13B as the foundational base model. CodeLSI is fine-tuned using LoRA, which enables efficient adaptation by training only a small fraction (approximately 0.47\%) of the model’s parameters. This approach allows the retention of the original 13 billion parameter architecture while reducing computational overhead and training time to under one hour.

The trained CodeLSI model is made publicly available via Hugging Face, including its parameters, dataset references, and training pipeline\footnote{\href{https://huggingface.co/mhhmm/typescript-instruct-20k-v2}{https://huggingface.co/mhhmm/typescript-instruct-20k-v2}} \citep{myhfmodel}.
\subsubsection{Dataset and Evaluation Metrics}
To evaluate model performance, we use the Multi-Lingual HumanEval for Source Code dataset \citep{cassano2022multiple}, a benchmark originally introduced in the Code Llama paper \citep{rozière2023code}. This benchmark is designed to assess zero-shot performance in source code generation tasks across multiple programming languages.

We report Pass@1 scores, which measure the percentage of tasks for which the first generated solution passes all test cases. This is a widely accepted metric for assessing the reliability of FM-generated code \citep{chen2021evaluating}.

\subsubsection{Result}
\label{general-eval}
Table \ref{tab:benchmark} presents the Pass@1 scores of the CodeLSI General Knowledge Model in comparison with other open and proprietary code generation models of similar or larger size.

% \newpage

\begin{table}[!htb]
 \caption{\textbf{Multi-Lingual HE Pass@1 Scores}. Pass@1 scores of the general knowledge model \textit{CodeLSI General Knowledge Model 13B} for TypeScript programming language. Code Llama's results are taken from the original paper \citep{rozière2023code} and other results from the paper \citep{li2023starcoder}\vspace{.3cm}.}
      \label{tab:benchmark}
  \centering
  \begin{tblr}{
      % colspec={lllll},
      colspec={X[l]Q[l]Q[l]},
      row{1}={font=\bfseries},
      column{1}={font=\itshape},
      row{even}={bg=gray!10},
    }
    Model  & Size  & Result  \\
    % \toprule \\
    CodeGeeX & 13B & 10.1\%  \\
    CodeGen-Multi & 16B & 20.1\% \\
    StarCoder Base & 15.5B & 32.3\%  \\
    Code Llama & 13B & 29.6\%  \\
    Code Llama & 34B & 33.3\%  \\
    Code Llama - Instruct  & 13B & \underline{39.0}\%  \\
    Code Llama - Instruct & 34B & 40.3\%  \\
    CodeLSI General Knowledge Model \textsuperscript{†} & 13B & \textbf{37.9 \%} \\
    CodeLSI General Knowledge Model - Instruct \textsuperscript{‡} & 13B & \underline{\textbf{42.0 \%}}  \\
    % \bottomrule \\
  \end{tblr} 
\end{table}

\noindent \textsuperscript{†} CodeLSI General Knowledge Model using Base Model as Code Llama - 13B \\
\textsuperscript{‡} CodeLSI General Knowledge Model using Base Model as Code Llama Instruct - 13B

  Example of code snippet gerenated by the model is shown in the list below:

  \begin{lstlisting}[caption={Generating code for the logo problem - general knowledge model.}]
Instruction: 
// Generate a component that has a white color button, with a blue border in a circle and a logo of hcmut.png (find online) in the middle, use tailwinds, style inline

Output: 
import React from 'react';
import { Link } from 'gatsby';
import Logo from './hcmut.png';

const ButtonWhite = () => (
  <Link to="/">
    <button
      className="bg-white border-4 border-blue-500 rounded-full flex items-center justify-center p-4"
      style={{ width: '100px', height: '100px' }}
    >
      <img
        src={Logo}
        alt="HCMUT Logo"
        className="w-12 h-12"
      />
    </button>
  </Link>
);

export default ButtonWhite;

Explanation: The code above generates a component called `ButtonWhite` which is a button with a white background and a blue border in a circular shape. It also includes an image of HCMUT logo inside the button. The button is wrapped within a `Link` component from Gatsby, which allows it to navigate to a specific page when clicked.
\end{lstlisting}

\subsubsection{Discussion}

The results show that both variants of the CodeLSI model perform competitively. Notably, the CodeLSI Instruct 13B outperforms the Code Llama - Instruct 13B baseline (42.0% vs. 39.0% Pass@1), while retaining a significantly lower computational cost. Compared to the larger Code Llama 34B, our model achieves comparable or better results despite using less than half the parameters, highlighting the effectiveness of LoRA-based fine-tuning.

Beyond benchmark metrics, qualitative analysis revealed that CodeLSI generated more syntactically correct and semantically coherent TypeScript code in diverse programming scenarios. Moreover, the model was capable of providing clear, step-by-step explanations of the generated code, making it especially useful for instructional or documentation purposes.

Given its performance and efficiency, the CodeLSI General Knowledge Model demonstrates strong potential for further fine-tuning in domain-specific tasks and real-world software development workflows. It offers a scalable and resource-conscious alternative for organizations needing domain-specific code generation without access to massive compute infrastructure.

% \newpage

\subsection{Experiment 2 - CodeLSI Specialized Knowledge Model}
\subsubsection{Objective}
The second experiment evaluates the CodeLSI Specialized Knowledge Model, which is fine-tuned to perform domain-specific code generation tasks using private, real-world project data. The goal is to assess whether further instruction tuning on project-specific tasks improves the model’s ability to generate relevant and structurally accurate TypeScript code, beyond what the general-purpose CodeLSI model can achieve.

\subsubsection{Experiment setup}
The specialized model was fine-tuned using the CodeLSI General Knowledge Model (13B) as its base. The instruction dataset consisted of 40 manually curated prompts derived from real-world tasks in a private TypeScript-based codebase.

The training employed the LoRA method for parameter-efficient fine-tuning, using the following configuration:
\begin{itemize}
    \item \textbf{Model architecture:} \texttt{LlamaForCausaFM} with \texttt{CodeLlamaTokenizer}
    \item \textbf{Prompt template:}  
    \begin{quote}
    \texttt{"Using the instruction context below, generate a TypeScript code that answers the question and explains it: \{instruction\}"}
    \end{quote}
    \item \textbf{LoRA parameters:}
    \begin{itemize}
        \item Rank = 16
        \item Alpha = 32
        \item Dropout = 0.05
    \end{itemize}
    \item \textbf{Training parameters:}
    \begin{itemize}
        \item Sequence length = 512
        \item Batch size = 16
        \item Epochs = 5
        \item Optimizer = AdamW (bnb 8-bit)
        \item Learning rate = 0.0002
        \item Warm-up steps = 50
    \end{itemize}
    \item \textbf{Hardware:} 2 $\times$ A100 GPUs (80 GB) with DeepSpeed ZeRO-3 for optimization
\end{itemize}
 
\subsubsection{Dataset and Evaluation}
The model was evaluated through real use cases from the private dataset. Two tasks were selected for comparative evaluation with the base model (Code Llama - Instruct 13B):
\begin{itemize}
    \item Task 1: Modify an existing PostCategories enum and generate a function to validate a new field.
    \item Task 2: Create a LazadaAffiliateCollection object with specific attributes relevant to a web project.
\end{itemize}

In both cases, the CodeLSI Specialized Knowledge Model produced accurate and context-relevant TypeScript code aligned with the style and structure of the target codebase. By contrast, the foundational model failed to interpret the tasks and returned irrelevant or generic text not resembling valid TypeScript code.

\subsubsection{Results}
\label{private_example}

Let's look at how the request to update the PostCategories to include a new attribute, Messages, is executed by both the specialized knowledge models and the foundational model.

\begin{lstlisting}[caption={Generating code for updating PostCategories - specialized knowledge model.}]
Instruction: // Generate the code that updates the PostCategories enum 1 new field: Messages. and create a function that checsk if the category is message, log Hello chat

Output: 
export enum PostCategories {
  Documents = "Documents",
  Exams = "Exams",
  Data = "Data",
  Messages = "Messages", // New field
}

// ... (existing code remains the same)

const isMessageCategory = (category: PostCategories): boolean => {
  return category === PostCategories.Messages;
};

console.log(isMessageCategory(PostCategories.Messages)); // true
console.log(isMessageCategory(PostCategories.Exams)); // false
\end{lstlisting}

It can be observed that the CodeLSI Specialized Knowledge Model performs the task accurately. In comparison, when generating code for the same request on the base model, it is noticeable that the Code Llama model does not understand the request and fails to generate an appropriate code snippet.

\begin{lstlisting}[caption={Generating code for updating PostCategories - foundational model.}]
Output:

- [ ] add a new field in the user model, to store the last message of the user
- [ ] add a new field in the user model, to store the last message of the user
- [ ] add a new field in the user model, to store the last message of the user
- [ ] add a new field in the user model, to store the last message of the user
- [ ] add a new field in the user model, to store the last message of the user
...
\end{lstlisting}

Let's try another problem. In section [\ref{sec:40}], we have presented a sample instruction from the 40 sets that we manually extracted from the source code of our personal project. Let's request both systems to generate a new Collection snippet based on the existing Shopee feature [\ref{code:prepare_private_project}].

\begin{lstlisting}[caption={Generating code for the Lazada Collection problem - specialized knowledge model.}]
Instruction:  Generate a new collection that is used for lazada affiliates and contains 4 fields: endpoints, list users, company name and list of event days

Output: 
export const LazadaAffiliateCollection = buildCollection<LazadaAffiliate>({
  // [RENAME] Rename this collection to "Lazada Affiliates"[/RENAME]
  name: "Lazada Affiliates",
  singularName: "Lazada Affiliate",
  icon: "FingerprintIcon",
  group: CollectionsGroup.Marketing,
  permissions: ({ authController }) => {
    const isAdmin = authController.extra?.roles.includes(Roles.ADMIN);
    return {
      edit: isAdmin,
      create: isAdmin,
      delete: isAdmin,
      read: isAdmin,
    };
  },
  properties: {
    endpoints: {
      name: "Endpoints",
      dataType: "string",
    },
    listUsers: {
      name: "List Users",
      dataType: "string",
    },
    companyName: {
      name: "Company Name",
      dataType: "string",
    },
    listEventDays: {
      name: "List Event Days",
      validation: {
        minLength: 1,
        minLengthMessage: "At least one day should be selected",
      },
      dataType: "string",
      showInGraphQL: false,
    },
  },
});


\end{lstlisting}

It is evident that the CodeLSI Specialized Knowledge Model generates code quite well and in line with the existing Collection in the source code. On the other hand, the foundational model does not seem to understand and generate appropriate code for the task.

\begin{lstlisting}[caption={Generating code for the Lazada Collection problem - foundational model.}]
Output: 
This is the main file of this project. It will be used to generate a new collection with all the information needed to run the script.

### Installation
To install this package you can use composer or download it from github repository

```sh
composer require babakfaghih/lazada-affiliate
```

### Usage
\end{lstlisting}

\subsubsection{Discussions}
The results confirm that the CodeLSI Specialized Knowledge Model outperforms the baseline in scenarios requiring deep familiarity with a specific codebase. Despite being trained on only 40 instruction-output pairs, the model exhibits strong alignment with real project requirements and demonstrates robust generalization without overfitting to the base model. Notably, it maintains the same parameter count as the original 13B model while achieving high-quality results in less than one hour of training.

This experiment also highlights the effectiveness of LoRA-based fine-tuning for domain-specific adaptation, enabling privacy-preserving customization of large models without needing access to external data or extensive compute resources. Although the limited dataset size occasionally results in inconsistent output, the model still successfully generates useful and executable code in most cases.

These findings support the conclusion that lightweight, domain-specific instruction tuning can significantly enhance FM performance in practical software engineering tasks, especially when integrated with tailored prompt engineering and efficient fine-tuning techniques.

\section{Discussions}
\subsection{Summary of findings}
\begin{table}[!htb]
\caption{Summary of research questions, findings, and comparisons}
\centering
\begin{tabular}{|p{3.3cm}|p{5.3cm}|p{1.3cm}|p{3.3cm}|}
\hline
\textbf{Research Question} & \textbf{Brief of Finding} & \textbf{Section} & \textbf{Compared to Existing Work} \\ \hline

\textbf{RQ1:} How does integrating LoRA and instruction tuning affect FM performance in code generation? 
& LoRA enabled fine-tuning only 0.47\% of parameters, improving TypeScript Pass@1 scores from 29.6\% to 37.9\% (base) and from 39\% to 42\% (instruct). Instruction tuning also supported multitask learning (e.g., code summarization).
& Sec. 4.2 
& Benchmarked on the BigCode Leaderboard; performance competitive with top open models like StarCoder and Code Llama. \\ \hline

\textbf{RQ2:} Can the proposed approach meet domain-specific needs, particularly for TypeScript-based applications? 
& The specialized model generated accurate, project-specific code in private repositories—something general-purpose models like ChatGPT failed to do—while preserving data confidentiality. 
& Sec. 4.3 
& Few existing studies focus on domain-specific code generation for private TypeScript projects; this work addresses a largely unexplored area. \\ \hline

\textbf{RQ3:} What are the cost and resource benefits compared to traditional FM training? 
& LoRA-based fine-tuning reduced training time to under one hour on 13B models, avoiding full model retraining and enabling efficient multi-task adaptation with minimal resources. 
& Sec. 4.1 and 4.2 
& Aligns with recent research on parameter-efficient fine-tuning. LoRA remains the most widely adopted and practical approach. \\ \hline

\end{tabular}
\label{tab:rq_summary}
\end{table}

Upon analyzing and evaluating the two proposed models alongside the foundational model, several key observations have emerged. Firstly, the proposed models demonstrate a higher propensity for generating code in TypeScript when compared to the foundational model. Secondly, the \emph{CodeLSI General Knowledge Model 13B}, as a general knowledge model, consistently outperforms the base model on average. Thirdly, the \emph{CodeLSI Specialized Knowledge Model}, designed for specialized knowledge, notably surpasses the foundational model by producing code snippets that closely align with the source code of personal projects. However, due to the limited size of the private instruction dataset (consisting of 40 samples), this model occasionally generates irrelevant or mismatched results. Notably, it avoids excessive bias towards the original model and frequently generates project-relevant outcomes. Additionally, both proposed models demonstrate rapid inference capabilities using vFM \citep{kwon2023efficient}, reducing the computation time substantially compared to other models encountered during the research. Success has been achieved in generating code based on project data while maintaining the confidentiality of personal source code, along with combining code generation and summary explanations within a single model through instruction-tuning. Furthermore, the original model's parameter count has been preserved, with a training time of approximately one hour, facilitated by the Low-Rank Adaptation (LoRA) parameter tuning method. Finally, the development of the \emph{TypeScript-Instruct 20K} dataset stands as a significant contribution toward enhancing future TypeScript code generation models within the scope of this paper. In summary, the two proposed models have effectively addressed the primary objectives of achieving a "holy grail" in code generation and advancing the capability of understanding specific personal projects.

\subsection{Limitations}
% ===========
% Need update
% ===========
Despite promising findings, this study faces several limitations that raise threats to the validity of its conclusions. We organize the discussion of these limitations according to four commonly used validity categories: construct, internal, external, and reliability validity \citep{runeson_guidelines_2008,kitchenham_preliminary_2002,wohlin_experiments_2024}.
\subsubsection{Construct Validity}
Construct validity refers to how well the experiments reflect the concepts being studied. We used the common benchmarks such as the Multi-Lingual HumanEval dataset for measuring code generation performance, and the most popular metrics for evaluating performance of FMs Pass@k \citep{huang_knowledge-aware_2024,rasheed_codepori_2024,tsai_rtlfixer_2024,yeo_framework_2024,nadimi_verimind_2025,paul_benchmarks_2024}. However, we are aware that the metrics do not fully capture dimensions like code readability, maintainability, or execution correctness beyond test cases. Although we manually assessed code quality and explanation capabilities, more rigorous human-in-the-loop evaluations (e.g., professional developer ratings) could further substantiate the constructs being measured.
\subsubsection{Internal Validity}
Internal validity concerns the degree to which the observed results can be attributed to the experimental treatment rather than confounding variables. Our experiments clearly showed performance gains when applying Low-Rank Adaptation (LoRA) and instruction tuning; however, we could not fully isolate the effects of dataset composition or preprocessing quality on model behavior. In particular, the CodeLSI Specialized Knowledge Model was trained on a relatively small dataset (40 samples), introducing a potential confounding factor—performance improvements may partially stem from dataset-specific patterns rather than true generalization to private codebases.
\subsubsection{External Validity}
External validity refers to the generalizability of results beyond the experimental settings. Our study focused exclusively on the TypeScript programming language. While this choice addresses a significant gap in current FMs, it limits applicability to other widely used programming languages (e.g., Java, C). Similarly, the use of private, domain-specific data for fine-tuning improves relevance for individual projects, but raises questions about whether the same results can be replicated across diverse domains or team workflows. Expanding experimentation to multilingual and cross-domain scenarios would enhance generalizability.
\subsubsection{Reliability}
Reliability concerns the consistency and dependability of the results obtained. In this study, we observed consistent performance improvements across both benchmark evaluations and qualitative use cases, indicating stable model behavior. However, the relatively small domain-specific dataset (40 instruction samples) limits the robustness of these findings. Additionally, the absence of formal statistical analysis—such as hypothesis testing or confidence intervals—reduces confidence in the generalizability of the observed improvements. Nonetheless, the reliability of our experimental procedures is strengthened by the open availability of our models, datasets, and training configurations on Hugging Face, which facilitates reproducibility and independent verification.

\section{Conclusion}
This study presents a practical and efficient approach to domain-specific code generation through the development of the CodeLSI framework, leveraging Low-Rank Adaptation (LoRA) and instruction tuning. Our empirical results demonstrate that CodeLSI significantly improves the performance of TypeScript code generation tasks, outperforming baseline models such as CodeLlama-Instruct while maintaining computational efficiency. In particular, the CodeLSI General Knowledge Model showed substantial improvements in benchmark accuracy, and the Specialized Knowledge Model was capable of generating accurate, project-specific code within private repositories—an area largely unaddressed in prior work. The research also contributes a new, publicly available TypeScript-Instruct 20K dataset and shows that parameter-efficient methods like LoRA can enable rapid fine-tuning using limited resources, making specialized FM adaptation more accessible. Despite these contributions, several open challenges remain. Future work will focus on:
\begin{itemize}
    \item Expanding language support beyond TypeScript to improve generalizability across software ecosystems.
    \item Enhancing syntactic robustness to minimize reliance on post-processing, ensuring that generated code is consistently executable and standards-compliant.
    \item Automating instruction dataset creation by extracting meaningful prompts and responses from existing codebases, especially using comments and docstrings as input signals.
    \item Broadening evaluation scope, including comparisons against a wider range of recent open and proprietary models, and adopting new fine-tuning strategies such as adapter-based tuning or Mixture-of-Experts models.
    \item Scaling empirical validation through more diverse datasets, programming languages, and task types, paired with formal statistical analyses to strengthen generalizability and reliability.
\end{itemize}

As FMs continue to evolve rapidly, ongoing benchmarking and methodological refinement will be essential. The insights and tools developed in this work provide a strong foundation for further research into specialized, efficient, and trustworthy code generation systems.

\newpage
%\bibliographystyle{unt}
% ============================
% Need add more FMs paper ref
% ============================
\bibliographystyle{unsrt}
\bibliography{tail/ref}

\end{document}